\documentclass[twoside]{dis07}
\usepackage[latin1]{inputenc}
\usepackage[dvips]{graphicx,epsfig,color}
\usepackage{wrapfig,rotating}
\usepackage{amssymb,amsmath,array}

\begin{document}
\title{Are generalized and transverse momentum dependent parton distributions
 related ?}

%***********************************************************************
% AUTHORS INFORMATION AREA
%***********************************************************************
\author{Stephan Mei{\ss}ner, Andreas Metz, and Klaus Goeke 
%
% Optional short acknowledgment: remove next line if non-needed
%
% DO NOT MODIFY THE FOLLOWING '\vspace' ARGUMENT
\vspace{.3cm}\\
%
% Addresses and institutions (remove "1- " in case of a single institution)
Institut f\"ur Theoretische Physik II, Ruhr-Universit\"at Bochum, \\
D-44780 Bochum, Germany
}

%***********************************************************************
% END OF AUTHORS INFORMATION AREA
%***********************************************************************

\maketitle

\begin{abstract}
The present knowledge on non-trivial relations between generalized parton 
distributions on the one hand and transverse momentum dependent distributions on 
the other is reviewed.
While various relations can be found in the framework of spectator models, 
so far no model-independent non-trivial relations have been established.
\end{abstract}

%%%%%%%%%%%%%%%%%%%%%%%%%%%%%%%%%%%%%%%%%%%%%%%%%%%%%%%%%%%%%%%%%%%%%%
\section{Definitions and trivial relations}
During the last decade a lot of effort has been devoted to study in detail 
generalized parton distributions (GPDs) as well as transverse momentum 
dependent parton distributions (TMDs).
While GPDs enter the QCD description of hard exclusive reactions on the 
nucleon, TMDs appear in connection with various semi-inclusive 
processes.
Recent work has suggested for the first time very interesting non-trivial relations 
between these two types of parton 
distributions~\cite{burkardt_02,burkardt_03a,burkardt_03b,diehl_05,burkardt_05,lu_06,meissner_07}.
The present short note~\cite{url} is based on Ref.~\cite{meissner_07}, where the 
current knowledge on such relations has been reviewed and previous work on this 
topic has been considerably extended.

To be specific now, two leading twist quark GPDs of the nucleon are defined 
through
\begin{eqnarray} \label{e:gpdcorr}
F^{q}(x,\Delta;\lambda,\lambda') & = &
 \frac{1}{2}\int\frac{dz^-}{2\pi}\,e^{i k\cdot z}\,
 \big<p';\lambda'\big|\,
 \bar\psi\big(\!-\!\tfrac{1}{2}z\big)\,\gamma^+ \,\mathcal{W}_{\textrm{GPD}} \,
 \psi\big(\tfrac{1}{2}z\big)\,\big|p;\lambda\big>\,
 \Big|_{z^+ = \vec{z}_T = 0} 
\nonumber \\
 & = & \frac{1}{2P^+}\,\bar u(p',\lambda')\,
 \bigg(\gamma^+\,H^{q}(x,\xi,t)+\frac{i\sigma^{+\mu}\Delta_\mu}{2M}\,
 E^{q}(x,\xi,t)\bigg)\, u(p,\lambda) \,, 
\end{eqnarray}
with $P = (p+p')/2$ denoting the average nucleon momentum and $\Delta=p'-p$
the momentum transfer to the nucleon. 
The GPDs $H^q$ and $E^q$ depend on the variables
\begin{equation}
 x=\frac{k^+}{P^+} \,, \qquad
 \xi=-\frac{\Delta^+}{2P^+} \,, \qquad
 t=\Delta^2 \,,
\end{equation}
where the dependence on the renormalization scale has been suppressed.
Note that the Wilson line $\mathcal{W}_{\textrm{GPD}}$ ensures the color gauge invariance 
of the bilocal quark operator in~(\ref{e:gpdcorr}).
The remaining six leading quark GPDs are obtained if one replaces the matrix 
$\gamma^+$ in the operator in~(\ref{e:gpdcorr}) by $\gamma^+ \gamma_5$ or 
$i\sigma^{j+}\gamma_5$ ($j$ being a transverse index).

In a similar way, two leading twist quark TMDs are defined according to
\begin{eqnarray} 
\Phi^{q}(x,\vec{k}_T;S) & = &
 \frac{1}{2}\int\frac{dz^-}{2\pi}\,\frac{d^2\vec{z}_T}{(2\pi)^2}\,e^{i k\cdot z}\,
 \big<P;S\big|\,\bar\psi\big(\!-\!\tfrac{1}{2}z\big)\,\gamma^+ \,
 \mathcal{W}_{\textrm{TMD}}\,\psi\big(\tfrac{1}{2}z\big)\,\big|P;S\big>\,\Big|_{z^+=0} 
\nonumber \\
 & = & f_1^{q}(x,\vec{k}_T^{\,2})
 -\frac{\epsilon^{ij}_T k_T^i S_T^j}{M}\,f_{1T}^{\bot q}(x,\vec{k}_T^{\,2}) \,.
\end{eqnarray}
The TMDs depend both on the longitudinal momentum fraction $x$ of the partons 
and on the transverse parton momentum $\vec{k}_T$.
While $f_1$ is the familiar unpolarized quark distribution, $f_{1T}^{\perp}$ 
represents the so-called Sivers function~\cite{sivers_89,boer_97}, which appears 
for a transversely polarized target and is supposed to be at the origin of various 
observed single spin phenomena in hard semi-inclusive reactions.

There exist some trivial relations between GPDs and TMDs because of the connection
between GPDs (for $\xi = t = 0$) and TMDs (integrated upon $\vec{k}_T$) on the one 
hand and ordinary parton distributions on the other.
An example is given by
\begin{equation}
H^q(x,0,0) = f_1^q(x) = \int d^2\vec{k}_T \, f_1^q(x,\vec{k}_T^2)\,. 
\end{equation}
Two additional trivial relations hold on the quark sector (involving the quark 
helicity and transversity distribution) and also two for gluon distributions.
In this note, however, we are mainly interested in non-trivial relations between
GPDs and TMDs.

%%%%%%%%%%%%%%%%%%%%%%%%%%%%%%%%%%%%%%%%%%%%%%%%%%%%%%%%%%%%%%%%%%%%%%
\section{Impact parameter representation of GPDs}
In Ref.~\cite{burkardt_02}, a non-trivial relation was proposed for the first time
--- a connection between the GPD $E$ and the Sivers function $f_{1T}^\perp$.
In that work an important role is played by the impact parameter representation
of GPDs.
For $\xi=0$, GPDs in impact parameter space have a density interpretation, and 
are generically given by
\begin{equation}
 \mathcal{X}(x,\vec{b}_T^{\,2})=\int\frac{d^2\vec{\Delta}_T}{(2\pi)^2} \,
 e^{-i\vec{\Delta}_T\cdot\vec{b}_T}\,X(x,0,-\vec{\Delta}_T^2) \,.
\end{equation}
Using this definition, the Fourier transform of the correlator in~(\ref{e:gpdcorr}) 
(for $\xi = 0$) has the form
\begin{equation} \label{e:impact}
\mathcal{F}^q(x,\vec{b}_T;S)
 = \int\frac{d^2\vec{\Delta}_T}{(2\pi)^2} \, 
   e^{-i\vec{\Delta}_T\cdot\vec{b}_T}\,F^q(x,\Delta_T;S) 
 = \mathcal{H}^{q}(x,\vec{b}_T^{\,2})
  +\frac{\epsilon_T^{ij}b_T^iS_T^j}{M}\,\bigg(\mathcal{E}^{q}(x,\vec{b}_T^{\,2})\bigg)' \,, 
\end{equation}
where the derivative of $\mathcal{E}^q$ with respect to $\vec{b}_T^{\,2}$ enters.
The correlator $\mathcal{F}^q$ has the following interpretation: it describes the 
distribution of unpolarized quarks carrying the longitudinal momentum fraction $x$ 
at a transverse position $\vec{b}_T$ inside a transversely polarized target.

If the second term on the {\it r.h.s.} in~(\ref{e:impact}) is non-zero, $\mathcal{F}^q$ 
is not axially symmetric in $b$-space.
In other words, the correlator is distorted.
In fact, one can show in a model-independent way that for a nucleon target the correlator 
has a large distortion, where the effect for a quark flavor $q$ is proportional to the 
contribution of the corresponding flavor to the anomalous magnetic moment of the 
nucleon~\cite{burkardt_02}.
One may now speculate that this large distortion should have an observable effect.  
Indeed in~\cite{burkardt_02} it was argued that it may be related to the Sivers 
function. 
An explicit form of the relation was obtained in Ref.~\cite{burkardt_03b} by considering 
the average transverse momentum of an unpolarized quark inside a transversely polarized 
target,
\begin{eqnarray} \label{e:rel} 
\big<k_T^{q,i}(x)\big>_{UT} & = &
 - \int d^2\vec{k}_T\,k_T^i\,\frac{\epsilon_T^{jk} k_T^j S_T^k}{M} \, 
   f_{1T}^{\bot q}(x,\vec{k}_T^{\,2})
 \nonumber \\
 & = & \int d^2\vec{b}_T\,\mathcal{I}^{q,i}(x,\vec{b}_T)\,
  \frac{\epsilon_T^{jk} b_T^j S_T^k}{M}\,
  \bigg(\mathcal{E}^q(x,\vec{b}_T^{\,2})\bigg)' \,.
\end{eqnarray}
The result in~(\ref{e:rel}) represents the first quantitative non-trivial relation 
between a GPD and a TMD.
It also provides an intuitive explanation of the Sivers effect.
(In this context we refer to~\cite{burkardt_02,burkardt_03a,burkardt_03b} where also 
the meaning of the object $\mathcal{I}^q$ is discussed.)
However, the relation~(\ref{e:rel}) is model-dependent.
It was obtained in the framework of a simple spectator model of the nucleon, treated 
to lowest non-trivial order in perturbation theory~\cite{burkardt_03b}.
On the other hand, the relation~(\ref{e:rel}) is quite successful from a phenomenological 
point of view.
Therefore, it makes sense to look for additional non-trivial relations, even if they turn
out to be merely model-dependent.

%%%%%%%%%%%%%%%%%%%%%%%%%%%%%%%%%%%%%%%%%%%%%%%%%%%%%%%%%%%%%%%%%%%%%%
\section{Model-independent considerations}
To get some guidance for further possible non-trivial relations the structures in the 
GPD- and TMD-correlator can be compared~\cite{diehl_05,meissner_07}. 
This procedure was first used in the case of quark distributions~\cite{diehl_05}, and 
later on extended to the gluon sector~\cite{meissner_07}.
Besides the already mentioned trivial relations (called relations of first type 
in~\cite{meissner_07}), one finds the following list of non-trivial 
analogies/relations between GPDs and TMDs~\cite{meissner_07}:
\begin{itemize}
\item Relations of second type
\begin{align} \label{e:type2}
 f_{1T}^{\bot q/g}\leftrightarrow-\Big(\mathcal{E}^{q/g}\Big)' \,,
  &\qquad h_1^{\bot q}\leftrightarrow
  -\Big(\mathcal{E}_T^{q}+2\mathcal{\tilde H}_T^{q}\Big)' \,, \notag\\
 \Big(h_{1T}^g+\tfrac{\vec{k}_T^{\,2}}{2M^2}\,h_{1T}^{\bot g}\Big)
  &\leftrightarrow -2\Big(\mathcal{H}_T^g-\tfrac{\vec{b}_T^{\,2}}{M^2}\,
 \Delta_b\mathcal{\tilde H}_T^g\Big)' \,.
\end{align}
\item Relations of third type
\begin{equation} \label{e:type3}
 h_{1T}^{\bot q}\leftrightarrow 2\Big(\mathcal{\tilde H}_T^{q}\Big)'' \,,
  \qquad h_1^{\bot g}\leftrightarrow 2\Big(\mathcal{E}_T^g+2\mathcal{\tilde H}_T^g\Big)'' \,.
\end{equation}
\item Relation of fourth type
\begin{equation} \label{e:type4}
 h_{1T}^{\bot g}\leftrightarrow -4\Big(\mathcal{\tilde H}_T^g\Big)''' \,.
\end{equation}
\end{itemize}
To the best of our knowledge Eqs.~(\ref{e:type2})--(\ref{e:type4}) contain 
all possible non-trivial analogies/relations between leading twist GPDs and 
TMDs for quarks and gluons.
Moreover, the method of Refs.~\cite{diehl_05,meissner_07} only indicates which 
distributions may be related, but does not provide an explicit form of a relation.

%%%%%%%%%%%%%%%%%%%%%%%%%%%%%%%%%%%%%%%%%%%%%%%%%%%%%%%%%%%%%%%%%%%%%%
\section{Model results}
In Ref.~\cite{meissner_07} we have studied two spectator models in order to find 
explicit forms of possible non-trivial relations: first, a scalar diquark spectator 
model of the nucleon; second, a quark target model treated in perturbative QCD,
which also allows one to study relations between gluon distributions.
We found it convenient to work with GPDs in momentum rather than impact parameter
representation.
The relations presented in the following involve moments of GPDs and TMDs, which 
(also for non-integer $n$) are defined according to
\begin{eqnarray} 
X^{(n)}(x) & = & 
 \frac{1}{2M^2}\int d^2\vec{\Delta}_T\,\bigg(\frac{\vec{\Delta}_T^2}{2M^2}\bigg)^{n-1}\,
 X(x,0,-\tfrac{\vec{\Delta}_T^2}{(1-x)^2}) \,,
 \\
Y^{(n)}(x) & = &
 \int d^2\vec{k}_T\,\bigg(\frac{\vec{k}_T^{\,2}}{2M^2}\bigg)^n\,Y(x,\vec{k}_T^{\,2}) \,. 
\end{eqnarray}
Taking as example the relation between the Sivers function and the GPD $E$,
the relations of the second type have the form~\cite{meissner_07}
\begin{equation} \label{e:modtype2}
f_{1T}^{\bot q\,(n)}(x) = h_2(n) \, \frac{1}{1-x} \,E^{q\,(n)}(x) \,, 
 \qquad \qquad (0 \le n \le 1) \,.
\end{equation}
The function $h_2(n)$ is different in the two models that we considered. 
We note that for all relations indicated in~(\ref{e:type2}) a formula 
corresponding to~(\ref{e:modtype2}) holds true.
Evaluating~(\ref{e:modtype2}) for $n=0$ and $n=1$ one recovers results presented
earlier in Refs.~\cite{lu_06,burkardt_03b}.
In this context it is also worthwhile to mention that for $n=1$ Eq.~(\ref{e:modtype2}) 
is equivalent to the content of Eq.~(\ref{e:rel}).

The model calculations provide the following explicit relation of third 
type~\cite{meissner_07},
\begin{equation} 
h_{1T}^{\bot q\,(n)}(x) = h_3(n) \, \frac{1}{(1-x)^2} \, \tilde{H}_T^{q\,(n)}(x) \,, 
 \qquad \qquad (0 \le n \le 1) \,,
\end{equation}
and a corresponding formula for the gluon distributions in~(\ref{e:type3}).
In contrast to the previous case the function $h_3$ is the same in both models.

Eventually, we mention that the relation of fourth type in~(\ref{e:type4}) is 
trivially satisfied in the quark target model, because to lowest non-trivial
order in perturbation theory both the TMD $h_{1T}^{\bot g}$ and the 
GPD $\tilde{H}_T^g$ vanish~\cite{meissner_07}.

%%%%%%%%%%%%%%%%%%%%%%%%%%%%%%%%%%%%%%%%%%%%%%%%%%%%%%%%%%%%%%%%%%%%%%
\section{Summary and discussion}
This note is dealing with the question if there exist non-trivial relations
between GPDs on the one hand and TMDs on the other.
On the basis of model-independent considerations one can distinguish between
different types of possible non-trivial relations.
It turns out that so far no model-independent non-trivial relations exist and 
it seems even unlikely that they can ever be established.
However, many relations exist in the framework of simple spectator models, 
treated to lowest non-trivial order in perturbation theory.
Once higher order diagrams are taken into consideration some of these relations 
are expected to break down~\cite{meissner_07}.
Nevertheless, for instance the phenomenology and the predictive power of the 
low-order spectator model relation between the Sivers effect and the GPD $E$ 
works quite well.
This is the only non-trivial relation which currently can be confronted with 
data.
Additional input from both the experimental and theoretical side is required 
in order to further study all other relations between GPDs and TMDs.
Future work will certainly shed more light on this interesting topic.

%%%%%%%%%%%%%%%%%%%%%%%%%%%%%%%%%%%%%%%%%%%%%%%%%%%%%%%%%%%%%%%%%%%%%%
\begin{footnotesize}

\end{footnotesize}

\end{document}